\documentclass[11pt,fleqn]{article} 
\usepackage{graphicx}
\usepackage{textcomp} 
\usepackage{mathtools}
\usepackage{epstopdf}
\usepackage{latexsym}
\usepackage{amsmath}
\usepackage{amssymb}
\usepackage{bm}
\usepackage[mathcal]{euscript}
\usepackage{slashed}
\usepackage{tikz}
\usepackage{MnSymbol}
\usepackage{cancel}

\usepackage[top=30mm, bottom=30mm, left=25mm, right=25mm]{geometry}

\numberwithin{equation}{section}

\usepackage{marvosym}

\usepackage{wasysym}

\usepackage{indentfirst}

\newcommand\blfootnote[1]{
  \begingroup
  \renewcommand\thefootnote{}\footnote{#1}
  \addtocounter{footnote}{-1}
  \endgroup
}

\usepackage{hyperref}
\hypersetup{
        unicode,	
        colorlinks,
        citecolor=blue,
        linkcolor=blue, 
        urlcolor=blue,
        bookmarksopen=true,
        bookmarksopenlevel=\maxdimen,
      }

\def\gl#1#2{\ifmmode \mathrm{GL}(#1; {\bf #2}) \else $\mathrm{GL}(#1; {\bf #2})$\fi}
\def\sl#1#2{\ifmmode \mathrm{SL}(#1; {\bf #2}) \else $\mathrm{SL}(#1; {\bf #2})$\fi}
\def\so#1{\ifmmode \mathrm{SO}({#1}) \else $\mathrm{SO}(#1)$\fi}

\def\sp#1#2{\ifmmode \mathrm{Sp}(#1; {\bf #2}) \else $\mathrm{Sp}(#1; {\bf #2})$\fi}
\def\usp#1{\ifmmode \mathrm{USp}(#1) \else $\mathrm{USp}(#1)$\fi}
\def\spin#1{\ifmmode \mathrm{Spin}(#1) \else $\mathrm{Spin}(#1)$\fi}
\def\su#1{\ifmmode \mathrm{SU}({#1}) \else $\mathrm{SU}(#1)$\fi}

\def\double #1{#1{\hbox{\kern-2pt $#1$}}}

\def\dd{\hbox{\,\Large$\triangleright$}}

\mathcode`\*="702A                  
\def\half{{\textstyle{1\over{\raise.1ex\hbox{$\scriptstyle{2}$}}}}}

\def \p{\partial}
\def \pb{\overline\partial}
\def \Jb{\overline J}
\def \Tb{\overline T}
\def \a{\alpha}
\def\ab{\overline\alpha}

\def \b{\beta}

\def\bb{\overline\beta}

\def \d{\delta}

\def \g{\gamma}

\def\gb{\overline\gamma}

\def \l{\lambda}

\def \lb{\overline\lambda}
\def \L{\Lambda}
\def \Lb{\overline\Lambda}

\def \ob{\bar\omega}
\def \O{\Omega}
\def\Oh{\widehat\Omega}
\def \t{\theta}

\def \N{\nabla}

\def \Nb{\overline\nabla}
\def\ua{\underline{a}}
\def\ub{\underline{b}}
\def\uc{\underline{c}}
\def\ud{\underline{d}}

\def \NN{\overline N}

\def \s{\sigma}

\def\Pib{\overline\Pi}

\def\psib{\overline\psi}
\def\dd{\overline d}
\def\wb{\overline w}

\def\Ch{\widehat C}
\def\Yh{\widehat Y}
\def\Dh{\widehat D}

\def\Vh{\widehat V}

\def\thb{\overline\theta}
\def\pp{\overline p}
\def\zb{\overline z}
\def\yb{\overline y}
\def\Wb{\overline W}
\def\cb{\overline c}

\def\hO{\widehat O}

\begin{document}

\begin{flushright}
\makebox[0pt][b]{}
\end{flushright}

\hspace{12cm} 

\vspace{40pt}
\begin{center}
{\LARGE \bf{A note on type II superstring vertex operators }}

\vspace{10pt}

{\LARGE \bf{in the B-RNS-GSS formalism}}

\vspace{40pt}
Osvaldo Chandia${}^{\star}$
\vspace{20pt}

{\em Departamento de Ciencias, Facultad de Artes Liberales
}\\
{\em Universidad Adolfo Ib\'a\~nez, Chile}\\



\vspace{60pt}
{\bf Abstract}
\end{center}
We construct integrated and unintegrated vertex operators for the type II superstring using the B-RNS-GSS formalism. The construction is done in flat spacetime background for both type II superstrings and type IIB superstring in a $AdS_5\times S^5$ background.

\blfootnote{
${}^{\star}$ \href{mailto:ochandiaq@gmail.com}{ochandiaq@gmail.com}}


\setcounter{page}0
\thispagestyle{empty}
\newpage

\tableofcontents

\parskip = 0.1in

\section{Introduction}

The Ramond-Neveu-Schwarz (RNS) superstring exhibits supersymmetry on the world-sheet, but its spacetime supersymmetric properties are difficult to verify \cite{Friedan:1985ge}. On the other hand, the Green-Scwharz-Siegel (GSS) supertstring is manifestly supersymmetric but it has not been possible to quantize it while preserving the symmetries of the spacetime background \cite{Siegel:1985xj}. About twenty-five years ago, Berkovits proposed a new superstring which has manisfest spacetime supersymmetry and can be quantized \cite{Berkovits:2000fe} and it is known as the pure spinor superstring (B). This formalism can be defined in any on-shell background spacetime even those with Ramond-Ramond background fields such as the classical example of type IIB superstring in a $AdS_5\times S^5$ background \cite{Berkovits:2000yr}. This new formalism is conformal invariant in flat and curved backgrounds \cite{Chandia:2003hn,Bedoya:2006ic}  but does not possess world-sheet supersymmetry. A new formalism was propesed that mixes all the good properties of the RNS, GSS and the B superstrings in \cite{Berkovits:2021xwh} and is called B-RNS-GSS formalism. 

This new formalism has a BRST charge that allows quantization and has the structure of a $N=2$ world-sheet supersymmetric BRST charge. Imposing this fact on a curved background it is possible to define heterotic \cite{Berkovits:2022dbm} and type II supertstrings \cite{Chandia:2023eel} .

The purpose of this paper is to construct massless vertex operators using the B-RNS-GSS formalism. Vertex operators describe physical states and are used in the computation of scattering amplitudes that we are not going to explore here. The structure is in the bosonic string that we review now. 

Consider the open bosonic string case. Physical states are defined in the cohomology of the BRST charge which, in the conformal gauge, is given by
\begin{align}
    Q=\oint dz\left(cT+bc\p c\right),
\end{align}
where $T=-\frac12\p X_m\p X^m$ is the stress-energy tensor for the matter $X$ world-sheet variables and $(b,c)$ are the Faddeev-Popov ghosts. 

The unintegrated vertex operator is defined as a zero conformal dimension and ghost number $1$ operator in the cohomology of $Q$, that is, $QU=0$ and $U\sim U+Q\L$. For the massless state, it is given by
\begin{align}\label{UObos}
    U=c\p X^m A_m(X),
\end{align}
which is physical if the field $A_m(X)$ satisfies the equations of a Maxwell field. The integrated vertex operator is a conformal weight operator $+1$ with vanishing ghost number. For the Maxwell field, the integrated vertex is given by
\begin{align}\label{VObos}
    V=\p X^m A_m(x).
\end{align}
Note that $U$ and $V$ are related by
\begin{align}\label{VUbos}
    QV=\p U.
\end{align}

Consider now the closed string. The BRST charge given by
\begin{align}
    Q=\oint dz\left(cT+bc\p c\right)+\oint d\overline{z}\left(\widetilde c \overline{T} + \widetilde b \widetilde c \pb \widetilde c\right) ,
\end{align}
where $T=-\frac12\p X_m\p X^m$ is the left-moving stress-energy tensor for the matter $X$ world-sheet variables, similarly $\overline T=-\frac12\pb X_m\pb X^m$ is the right-moving stress-energy tensor for the matter $X$ world-sheet variables. There are left-moving ghosts $(b,c)$ and right-moving ghosts $(\widetilde b,\widetilde c)$.

The unintegrated vertex operator for the graviton is a conformal dimension $(0,0)$ and ghost number $(1,1)$ state in the cohomology of the BRST and it is given by
\begin{align}\label{UCbos}
    U=c\widetilde c\p X^m \pb X^n h_{mn}(X),
\end{align}
which turns out to describe a physical state when the field $h_{mn}$ satisfies the linearized Einstein equations. The corresponding integrated vertex operator has conformal weight $(1,1)$ and vanishing ghost number. It is given by
\begin{align}\label{VCbos}
    V=\p X^m \pb X^n h_{mn}(X).
\end{align}
The relations between $U$ and $V$ involves the operators $W$ and $\overline{W}$ which satisfy
\begin{align}\label{VUCbos}
    QV=\pb W-\p\overline{W},\quad QW=\p U,\quad Q\overline{W}=\pb U. 
\end{align}
The operators $W$ and $\overline{W}$ are given by
\begin{align}
    W=\widetilde c\p X^m \pb X^n h_{mn}(X),\quad \overline{W}=-c\p X^m \pb X^n h_{mn}(X).
\end{align}

Note that we could start with the most general ghost number $(1,1)$ and conformal weight $(0,0)$ representing the uninintegrated vertex operator $U$ and then derive $W, \Wb$ and $V$. Alternatively, one could start with the most general conformal weight $(1,1)$ and vanishing ghost number representing the integrated vertex operator $V$ and then derive $W, \Wb$ and $U$. I chose the second possibility because it is easier to figure out what is this most general $V$ and without ghost dependence. The idea of this paper is to establish a relation like (\ref{VUCbos}) for the massless states in the type II superstring using the B-RNS-GSS formalism in flat spacetime and in $AdS_5\times S^5$ background starting with the most general integrated vertex operator. In section 2 we consider a superstring in flat spacetime and review the construction of massless vertex operators in the open superstring case and then we construct integrated and unintegrated vertex operators for the closed superstring. In section 3 we study the special case of the type IIB superstring in $AdS_5\times S^5$ background and construct integrated and unintegrated vertex operators.

\section{The flat spacetime case}
The B-RNS-GSS formalism was introduced in \cite{Berkovits:2021xwh} and we review the open string case (or the left-moving sector of the heterotic string) and below we will consider the type II case. The world-sheet action and the BRST charge are obtained by performing a similarity transformation on the RNS system together with the addition of the  free field conjugate pairs $(\theta^\alpha,p_\a)$ and $(\L^\a,w_\a)$. Here $\a=1,\dots 16$, $\theta^\a$ are the $N=1$ superspace coordinates in ten dimensions. Recall that the RNS world-sheet fields are the spacetime coordinates $X^m$ and their supersymmetry partners $\psi^m$ together with super-reparametrization ghosts $(b, c,  \b, \g)$. The action is given by
\begin{align}
    S=\int d^2z ~\left( \frac12 \p X_m \pb X^m+\frac12 \psi_m\pb\psi^m + b \pb c + \b \pb \g + p_\a \pb \theta^\a + w_\a \pb \L^\a \right),
\end{align}
where the unconstrained bosonic variables $(w,\L)$ are in the untwisted case so they have conformal weights $(\frac12,\frac12)$, respectively. The BRST charge is given by
\begin{align}
    &Q=\oint dz \left( cT-bc\p c+b\g^2+\g G\right),\cr 
    &G=\L^\a d_\a+\psi^m\Pi_m+w_\a\p\theta^\a+\frac12(\L\g_m\L)\psi^m
\end{align}
where $T$ is the stress-energy tensor given by
\begin{align}
    T=-\frac12\Pi_m\Pi^m-d_\a\p\theta^\a-\frac12\psi_m\p\psi^m- \frac12 w_\a\p\L^\a-\frac12\L^\a\p w_\a-\frac32\b\p\g-\frac12\g\p\b-2b\p c+ c\p b ,
\end{align}
and
\begin{align}
    &\Pi^m=\p X^m+\frac12(\theta\g^m\p\theta),\cr
    &d_\a=p_\a-\frac12(\g^m\theta)_\a \p X_m -\frac18(\g^m\theta)_\a(\theta\g_m\p\theta) .
\end{align}
Although to achieve nilpotency of the BRST charge one needs to add non-minimal bosonic variables as is discussed in \cite{Berkovits:2022dbm}, our construction of massless vertex operators does not require these non-minimal variables so they are ignored in this paper. 

According to the discussion of \cite{Berkovits:2022dbm}, physical states are not only required to be in the cohomology of the BRST charge, they also have to contain terms that carry non-positive charge with respect to 
\begin{align}\label{charge}
    J=\oint\left(\xi\eta -w_\a\L^\a\right),
\end{align}
where the $\xi$ and $\eta$ come from the fermionization of the RNS ghosts $(\b,\g)$ as $\b=\p\xi e^{-\phi}, \g=\eta e^\phi$. Note that $\L$ carries charge $+1$ and $w$ carries charge $-1$. It turns out that the integrated vertex operator $V$ for the massless states has terms of charge $0$ and $-1$ and is given by
\begin{align}\label{Vopen}
    V=\p\theta^\a A_\a+\Pi^m A_m+d_\a W^\a+\frac12 N^{mn}F_{mn}-\psi^m w_\a \p_m W^\a,
\end{align}
where $N^{mn}=-\psi^m\psi^n+\frac12(\L\g^{mn}w)$. The superfields $(A_\a, A_m, W^\a, F_{mn}=\p_{[m}A_{n]})$ are constrained as 
\begin{align}
    &D_\a A_\b+D_\b A_\a=\g^m_{\a\b} A_m ,\cr
    &D_\a A_m-\p_m A_\a=(\g_m)_{\a\b}W^\b,\cr 
    &\p_m A_n-\p_n A_m=F_{mn}.
\end{align}
It can be shown that 
\begin{align}
    \p^m F_{mn}=0,\quad \g^m_{\a\b}\p_m W^\b=0,
\end{align}
then the theta independent components of $A_m$ and $W^\a$ describe the photon and the photino.  We would like to have a relation like (\ref{VUbos}) so we obtain the unintegrated vertex operator. In the Lorentz gauge
\begin{align}
    \p^m A_m=0 ,
\end{align}
the integrated vertex oprerator $V$ is primary, that is
\begin{align}
    T(y)V(z)\to\frac{V(z)}{(y-z)^2}+\frac{\p V(z)}{(y-z)}.
\end{align}
We now prove that there exists an unintegrated vertex operator $U$ that satisfies $QV=\p U$, just like in the bosonic case of (\ref{VUbos}). First we note that
\begin{align}
    G(y) V(z)\to\frac{u(z)}{(y-z)^2}+\frac{\p u(z)}{(y-z)}, 
\end{align}
where $u$ is a conformal weight $\frac12$ world-sheet field an it is given by
\begin{align}
    u=\L^\a A_\a-\psi^m A_m-w_\a W^\a .
\end{align}
Note that $u$ has terms with charges $(1,0,-1)$ respect to (\ref{charge}). Finally, we obtain that the unintegrated vertex operator satisfying $QV=\p U$ is
\begin{align}
    U=cV+\g u.
\end{align}
Note that, in terms of the charges given by (\ref{charge}), the integrated vertex $V$ contains terms with less possible charges that the terms in the unintegrated $U$. We will use this fact to start with $V$ in closed string case.

Consider the type II case. The world-sheet action and the BRST charge are obtained by performing a similarity transformation on the RNS system together with the free field conjugate pairs $(\theta^\alpha,p_\a),(\overline\theta^{\overline\a},\overline{p}_{\overline\a})$ and $(\L^\a,w_\a),(\Lb^{\ab},\wb_{\ab})$. Here $\a=1,\dots 16$, $(\theta^\a,\thb^{\ab})$ are the $N=2$ superspace coordinates in ten dimensions. Recall that the RNS world-sheet fields are the space-time coordinates $X^m$ and their supersymmetry partners $(\psi^m,\psib^m)$ together with superreparametrization ghosts $(b,c,\b,\g),(\overline b,\overline c,\bb,\gb)$. The action is given by
\begin{align}
    S=\int d^2z ~&\left( \frac12 \p X_m \pb X^m+\frac12 \psi_m\pb\psi^m+\frac12\psib_m\p\psib^m + b \pb c+\overline b\p\overline c + \b \pb \g+\bb\p\gb \right.\cr
    &+\left. p_\a \pb \theta^\a+\pp_{\ab}\p\thb^{\ab} + w_\a \pb \L^\a+\wb_{\ab}\p\Lb^{\ab} \right),
\end{align}
where the unconstrained bosonic variables $(w,\L),(\wb,\Lb)$ are in untwisted case so that $(\l,w)$ have conformal weights $(\frac12,0)$ and $(\lb,\wb)$ have conformal weights $(0,\frac12)$. The BRST charges are given by
\begin{align}
    &Q_L=\oint dz \left( cT-bc\p c+b\g^2+\g G_L\right),\cr
    &G_L= \L^\a d_\a+\psi^m\Pi_m+w_\a\p\theta^\a+\frac{1}{2}(\L\g_m\L)\psi^m ,\cr  
    &Q_R=\oint d\overline z \left( \overline c\Tb-\overline b\overline c\pb\overline c+\overline b\gb^2+\gb G_R\right),\cr 
    &G_R=\Lb^{\ab} \dd_{\ab}+\psib^m\Pib_m+\wb_{\ab}\pb\thb^{\ab}+\frac{1}{2}(\Lb\g_m\Lb)\psib^m ,
\end{align}
where $(T,\Tb)$ are the stress-energy tensor components and are given by
\begin{align}
    &T=-\frac12\Pi_m\Pi^m-d_\a\p\theta^\a-\frac12\psi_m\p\psi^m-\frac12w_\a\p\L^\a-\frac12\L^\a\p w_\a-\frac32\b\p\g-\frac12\g\p\b-2b\p c+ c\p b ,\cr 
    &\Tb=-\frac12\Pib_m\Pib^m-\dd_{\ab}\pb\thb^{\ab}-\frac12\psib_m\pb\psib^m-\frac12\wb_{\ab}\pb\Lb^{\ab}-\frac12\Lb^{\ab}\pb\wb_{\ab}-\frac32\bb\pb\gb-\frac12\gb\pb\bb-2\overline b\pb\overline c+ \overline c\pb\overline b ,
\end{align}
and
\begin{align}
    &\Pi^m=\p X^m+\frac12(\theta\g^m\p\theta),\quad \Pib^m=\pb X^m+\frac12(\thb\g^m\pb\thb),\cr
    &d_\a=p_\a-\frac12(\g^m\theta)_\a \p X_m -\frac18(\g^m\theta)_\a(\theta\g_m\p\theta),\cr 
    &\dd_{\ab}=\pp_{\ab}-\frac12(\g^m\thb)_{\ab} \pb X_m -\frac18(\g^m\thb)_{\ab}(\thb\g_m\pb\thb).
\end{align}


The integrated vertex operator $V$ for the massless states can be obtained from the left-right product of $V_L\otimes V_R$ where $V_L$ and $V_R$ have the form of (\ref{Vopen}). The integrated vertex becomes
\begin{align}
    V=&\left[\p\theta^\a A_\a+\Pi^m A_m+d_\a W^\a+\frac12 N^{mn}F_{mn}-\psi^m w_\a \p_m W^\a\right]\otimes \cr 
    &\left[\pb\thb^{\ab} \widetilde A_{\ab}+\Pib^m \widetilde A_m+\dd_{\ab} \widetilde W^{\ab}+\frac12\NN^{mn}\widetilde F_{mn}-\psib^m \wb_{\ab} \p_m \widetilde W^{\ab}\right] \cr
    &\equiv\p\theta^\a\pb\thb^{\ab}A_{\a\ab}+\Pi^m\Pib^n A_{mn}+d_\a \dd_{\ab} P^{\a\ab}+ N^{mn}\NN^{pq} S_{mnpq}+\psi^m w_\a\psib^n \wb_{\ab} U_{mn}{}^{\a\ab} \cr 
    &+\p\theta^\a\Pib^m A_{\a m}+\Pi^m\pb\thb^{\ab} A_{m\ab}+\p\theta^\a \dd_{\ab} E_\a{}^{\ab}+d_\a\pb\thb^{\ab}E_{\ab}{}^\a \cr
    &+\p\theta^\a\NN^{mn}\Oh_{\a mn}+ N^{mn}\pb\thb^{\ab}\O_{\ab mn}+\p\theta^\a\psib^m\wb_{\ab}\Ch_{m\a}{}^{\ab}+\psi^m w_\a\pb\thb^{\ab} C_{m\ab}{}^\a \cr 
    &+\Pi^m \dd_{\ab} E_m{}^{\ab}+d_\a\Pib^m E_m{}^\a+\Pi^m \NN^{np} \Oh_{mnp}+N^{np}\Pib^m \O_{mnp} \cr 
    &+\Pi^m \psib^n\ob_{\ab} C_{mn}{}^{\ab}+\psi^n w_\a \Pib^m C_{mn}{}^\a+d_\a \NN^{mn} \Dh_{mn}{}^\a+N^{mn} \dd_{\ab} D_{mn}{}^{\ab}  \cr 
    &+d_\a\psib^m\wb_{\ab} V_m{}^{\a\ab}+\psi^m w_\a \dd_{\ab} \Vh_m{}^{\ab\a}+N^{mn} \psib^p\wb_{\ab}Y_{mnp}{}^{\ab}+\psi^p w_\a\NN^{mn}\Yh_{mnp}{}^\a . 
\end{align}
The superfields in $V$ satisfy the constraint equations 
\begin{align}
    &D_{(\a}A_{\b)\ab}=-\g^m_{\a\b}A_{m\ab},\quad D_\a A_{m\ab}+\p_m A_{\a\ab}=-(\g_m)_{\a\b}E_{\ab}{}^\b,\quad \O_{\ab mn}=\frac12\p_{[m}A_{n]\ab},\cr 
    &D_{(\a}A_{\b)m}=\g^n_{\a\b}A_{nm},\quad D_\a A_{nm}-\p_n A_{\a m}=(\g_n)_{\a\b}{}E_m{}^\b,\quad \O_{mnp}=\frac12\p_{[n}A_{p]m},\cr 
    &D_{(\a}E_{\b)}{}^{\ab}=-\g^m_{\a\b}E_m{}^{\ab},\quad D_{\a}E_m{}^{\ab}+\p_m E_{\a}{}^{\ab}=-(\g_m)_{\a\b}P^{\b\ab},\quad D_{mn}{}^{\ab}=\frac12\p_{[m}E_{n]}{}^{\ab},\cr
    &D_{(\a}\Oh_{\b)mn}=\g^p_{\a\b}\Oh_{pmn},\quad D_\a\Oh_{pmn}-\p_p\Oh_{\a mn}=(\g_p)_{\a\b}\Dh_{mn}{}^\b,\quad S_{pqmn}=\frac12\p_{[p}\Oh_{q]mn},
\end{align}
and
\begin{align}
    &D_{(\ab}A_{\a\bb)}=\g^m_{\ab\bb}A_{\a m},\quad D_{\ab}A_{\a m}-\p_m A_{\a\ab}=(\g_m)_{\ab\bb}E_\a{}^{\bb},\quad \Oh_{\a mn}=\frac12\p_{[m}A_{\a n]},\cr
    &D_{(\ab}A_{m\bb)}=\g^n_{\ab\bb}A_{mn},\quad D_{\ab}A_{mn}-\p_n A_{m\ab}=(\g_n)_{\ab\bb}E_m{}^{\bb},\quad \Oh_{mnp}=\frac12\p_{[n}A_{mp]},\cr 
    &D_{(\ab}E_{\bb)}{}^\a=\g^m_{\ab\bb}E_m{}^\a,\quad D_{\ab}E_m{}^\a-\p_m E_{\ab}{}^\a=(\g_m)_{\ab\bb}P^{\a\bb},\quad \Dh_{mn}{}^\a=\frac12\p_{[m}E_{n]}{}^\a,\cr
    &D_{(\ab}\O_{\bb)mn}=\g^p_{\ab\bb}\O_{pmn},\quad D_{\ab}\O_{pmn}-\p_p\O_{\ab mn}=(\g_p)_{\ab\bb}D_{mn}{}^{\bb},\quad S_{mnpq}=\frac12\p_{[p}\O_{q]mn},
\end{align}
and finally
\begin{align}
    &C_{m\ab}{}^\a+\p_m E_{\ab}{}^\a=C_{nm}{}^\a+\p_m E_n{}^\a=\Vh_m{}^{\ab\a}+\p_m P^{\a\ab}=\Yh_{npm}{}^\a+\p_m\Dh_{np}{}^\a=0,\cr 
    &\Ch_{m\a}{}^{\ab}+\p_m E_\a{}^{\ab}=C_{nm}{}^{\ab}+\p_m E_n{}^{\ab}=V_m{}^{\a\ab}+\p_m P^{\a\ab}=Y_{npm}{}^{\ab}+\p_m D_{np}{}^{\ab}=0,\cr 
    &U_{mn}{}^{\a\ab}-\p_m\p_n P^{\a\ab}=0.
\end{align}
As in the open string case, the gauge-fixing conditions
\begin{align}
    &\p^m A_{m\ab}=\p^m A_{mn}=\p^m E_m{}^{\ab}=\p^m\Oh_{mnp}=\p^m C_{mn}{}^{\ab}=0,\cr 
    &\p^mA_{\a m}=\p^m A_{nm}=\p^m E_m{}^\a=\p^m\O_{mnp}=\p^m C_{mn}{}^\a=0,
\end{align}
help to set the integrated vertex operator $V$ to be a primary world-sheet field. That is,
\begin{align}
    &T(y) V(z,\overline{z})\to\frac{V(z,\overline{z})}{(y-z)^2}+\frac{\p V(z,\overline{z})}{(y-z)},\cr
    &\Tb(\overline{y})V(z,\overline{z})\to\frac{V(z,\overline{z})}{(\overline{y}-\overline{z})^2}+\frac{\pb V(z,\overline{z})}{(\overline{y}-\overline{z})}.
\end{align}
Our purpose is to prove that we have vertex operators satisfying the equations like (\ref{VUCbos}), that is
\begin{align}\label{theEQ}
    QV=\pb W-\p\overline W ,\quad QW=\p U,\quad Q\overline{W}=\pb U .
\end{align}
Given the integrated vertex operator $V$ we first note that 
\begin{align}
    &G_L(y)V(z,\zb)\to\frac{w_L(z,\zb)}{(y-z)^2}+\frac{\p w_L(z,\zb)}{(y-z)},
\end{align}
\begin{align}
    &G_R(\yb)V(z,\zb)\to\frac{w_R(z,\zb)}{(\yb-\zb)^2}+\frac{\pb w_R(z,\zb)}{(\yb-\zb)},
\end{align}
where
\begin{align}
    &w_L=\pb\thb^{\ab}\left(\L^\a A_{\a\ab}+\psi^m A_{m\ab}-w_\a E_{\ab}{}^\a\right)+\Pib^m\left( \L^\a A_{\a m}-\psi^n A_{nm}-w_\a E_m{}^\a \right) \cr 
    &+\dd_{\ab}\left( \L^\a E_\a{}^{\ab}+\psi^m E_m{}^{\ab}-w_\a P^{\a\ab} \right)+\NN^{mn}\left( \L^\a\Oh_{\a mn}-\psi^p\Oh_{pmn}-w_\a\Dh_{mn}{}^\a \right)\cr 
    &+\psib^m \wb_{\ab}\left( \L^\a\Ch_{m\a}{}^{\ab}+\psi^n C_{nm}{}^{\ab}-w_\a V_m{}^{\a\ab} \right) ,
\end{align}
\begin{align}
    &w_R=\p\t^{\a}\left(-\Lb^{\ab} A_{\a\ab}+\psib^m A_{\a m}+\wb_{\ab} E_{\a}{}^{\ab}\right)+\Pi^m\left( \Lb^{\ab} A_{m\ab}-\psib^n A_{mn}-\wb_{\ab} E_m{}^{\ab} \right) \cr 
    &+d_{\a}\left(-\Lb^{\ab} E_{\ab}{}^{\a}+\psib^m E_m{}^{\a}+\wb_{\ab} P^{\a\ab} \right)+N^{mn}\left( \Lb^{\ab}\O_{\ab mn}-\psib^p\O_{pmn}-\wb_{\ab}D_{mn}{}^{\ab} \right)\cr 
    &+\psi^m w_{\a}\left(-\Lb^{\ab}C_{m\ab}{}^{\a}+\psib^n C_{nm}{}^{\a}+\wb_{\ab} \Vh_m{}^{\ab\a} \right) .
\end{align}
Using these OPE's, the BRST transformation of the integrated vertex operator $V$ becomes
\begin{align}
    QV=\pb W-\p\Wb, 
\end{align}
with
\begin{align}\label{Wflat}
    W=\cb V+\gb w_R,\quad \Wb=-cV-\g w_L .
\end{align}
In order to obtain the relations in (\ref{theEQ}) we prove now that
\begin{align}\label{QWs}
    QW=\p U,\quad Q\Wb=\pb U.
\end{align}
Since $G_L(y)G_L(z)\to -T(z)/(y-z)$ and $G_R(\yb)G_R(\zb)\to-\Tb/(\yb-\zb)$ we obtain
\begin{align}
    G_L(y)w_L(z,\zb)\to-\frac{V(z,\zb)}{(y-z)},\quad G_R(\yb)w_R(z,\zb)\to-\frac{V(z,\zb)}{(\yb-\zb)}.
\end{align}
To obtain $QW$ and $Q\Wb$ we also need $G_L(y)w_R(z,\zb)$ and $G_R(\yb)w_L(z,\zb)$. They are
\begin{align}
    &G_L(y)w_R(z,\zb)\to\frac{u(z,\zb)}{(y-z)^2}+\frac{\p u(z,\zb)}{(y-z)},\cr
    &G_R(\yb)w_L(z,\zb)\to-\frac{u(z,\zb)}{(\yb-\zb)^2}-\frac{\pb u(z,\zb)}{(\yb-\zb)},
\end{align}
where
\begin{align}
    &u=-\L^\a\Lb^{\ab}A_{\a\ab}+\L^\a\psib^m A_{\a m}-\psi^m\Lb^{\ab}A_{m\ab}+\L^\a\wb_{\ab}E_\a{}^{\ab}+w_\a\Lb^{\ab} E_{\ab}{}^\a\cr
    &+\psi^m\psib^n A_{mn}+\psi^m\wb_{\ab}E_m{}^{\ab}-w_\a\psib^m E_m{}^\a-w_\a\wb_{\ab}P^{\a\ab}.
\end{align}
Note that $w_L$ is a primary field of conformal weights $(\frac12,1)$ and $w_R$ is a primary field of conformal weights $(1,\frac12)$. Finally, applying the above formulae we obtain (\ref{QWs}) with the unintetegrated vertex operator being equal to
\begin{align}\label{Uflat}
    U=c\cb V+c\gb w_R-\cb\g w_L+\g\gb u.
\end{align}

In this section we have constructed the vertex operators that satisfy (\ref{theEQ}) for the type II superstring. In the next section the same goal is reached for the case of the type IIB superstring in $AdS_5\times S^5$ background.

\section{The $AdS_5\times S^5$ background case}

We now consider the type IIB superstring in $AdS_5\times S^5$ \cite{Chandia:2023eel}. The world-sheet action is given by 
\begin{align}\label{Sads}
    &S=\int d^2z~\left(\frac12 J_{\underline{a}}\Jb^{\underline{a}}+\frac12\eta_{\a\ab}(J^\a\Jb^{\ab}+J^{\ab}\Jb^\a)+d_\a\Jb^\a+\dd_{\ab}J^{\ab}-\frac12\eta^{\a\ab}d_\a\dd_{\ab}+w_\a\Nb\L^\a+\wb_{\ab}\N\Lb^{\ab}\right.\cr 
    &+\left.\frac12\psi_{\ua}\Nb\psi^{\ua}+\frac12\psib_{\ua}\N\psib^{\ua}-\frac12\psi^{\ua}w_\a\Jb^{\ab}(\eta\g_{\ua})^\a{}_{\ab}+\frac12\psib^{\ua}\wb_{\ab}J^\a(\g_{\ua}\eta)_\a{}^{\ab}-\frac18 \Jb^{\ua}(w\g_{\ua}w)\right.\cr
    &-\left.\frac18 J^{\ua}(\wb\g_{\ua}\wb)+\frac12 N^{[\underline{ab}]}\NN_{[\underline{ab}]}-\frac18\psi^{\ua}\psib^{\ub}w_\a\wb_{\ab}(\g_{\ua}\g_{\ub}\eta)^{\a\ab}-\frac1{64}(w\g_{\ua}w)(\wb\g^{\ua}\wb)\right)+S_{\mbox{ghosts}},
\end{align}
where $J=g^{-1}dg$ are the left-invariant currents of the superalgerbra $PSU(2,2|4)/SO(4,1)\times SO(5)$ which provides the supercoset description of the $AdS_5\times S^5$ background \cite{Berkovits:1999zq}. The background geometry is obtained from the Ramond-Ramond field strength $P^{\a\bb}=-\frac12\eta^{\a\bb}$ and the non-vanishing component of the Kalb-Ramond tensor $B_{\a\bb}=\eta_{\a\bb}$, where $\eta_{\a\bb}=(\g^0\g^1\g^2\g^3\g^4)_{\a\bb}$ and $\eta^{\a\bb}=(\g_0\g_1\g_2\g_3\g_4)^{\a\bb}$. We can also define
\begin{align}
    \g_{\ua}^{\a\b}=\eta^{\a\ab}\eta^{\b\bb}(\g_{\ua})_{\ab\bb},\quad \g_{\ua}^{\ab\bb}=\eta^{\a\ab}\eta^{\b\bb}(\g_{\ua})_{\a\b}.
\end{align}
The ten-dimensional vector index $\ua$ is split into the vector index $a$ for $AdS_5$ and the vector index $a'$ for $S^5$. The covariant derivatives in the action are defined by
\begin{align}
    &\Nb\L^\a=\pb\L^\a+\frac14\L^\b(\g^{\underline{ab}})_\b{}^\a \Jb_{\underline{ab}},\quad \Nb\psi^{\ua}=\pb\psi^{\ua}+\psi^{\underline{b}}\Jb_{\underline{b}}{}^{\ua},\cr 
    &\N\Lb^{\ab}=\p\Lb^{\ab}+\frac14\Lb^{\bb}(\g^{\underline{ab}})_{\bb}{}^{\ab} J_{\underline{ab}},\quad \N\psi^{\ua}=\p\psi^{\ua}+\psi^{\underline{b}} J_{\underline{b}}{}^{\ua}.\cr
\end{align}
For the terms with $N\NN$ we are using the notation $A^{[\underline{abc\dots}]}B_{[\underline{abc\dots}]}=A^{abc\dots}B_{abc\dots}-A^{a'b'c'\dots}B_{a'b'c'\dots}$. The last term in (\ref{Sads}) $S_{\mbox{ghosts}}$ is the action of the RNS ghosts. The BRST charge is given by $Q=Q_L+Q_R$, where
\begin{align}
    &Q_L=\oint dz\left(cT-bc\p c+b\g^2+\g G_L \right),\cr
    &G_L=\L^\a d_\a+\psi_{\ua}J^{\ua}+w_\a J^\a+\frac1{2}(\L\g_{\ua}\L)\psi^{\ua}-\frac18\psi^{\ua}(w \g_{\ua} w)
\end{align}
\begin{align}
    &Q_R=\oint d\overline z\left(\overline c\overline T-\overline b\overline c\pb \overline c+\overline b\gb^2+\gb G_R \right),\cr 
    &G_R=\Lb^{\ab} \dd_{\ab}+\psib_{\ua}\Jb^{\ua}+\wb_{\ab} \Jb^{\ab}+\frac1{2}(\Lb\g_{\ua}\Lb)\psib^{\ua}-\frac18\psib^{\ua}(\wb\g_{\ua}\wb),
\end{align}
where the stress-energy tensor components are given by
\begin{align}
    T&=-\frac12 J_{\ua} J^{\ua}-d_\a J^\a-\frac12 w_\a\N\L^\a-\frac12\L^\a\N w_\a-\frac12\psi_{\ua}\N\psi^{\ua}\cr
    &+\frac12\psi^{\ua}w_\a J^{\ab} (\eta\g_{\ua})^\a{}_{\ab}+\frac18 J^{\ua}(w\g_{\ua}w)+T_{\mbox{ghosts}} ,
\end{align}
\begin{align}
    \Tb&=-\frac12 \Jb_{\ua} \Jb^{\ua}-\dd_{\ab} \Jb^{\ab}-\frac12\wb_{\ab}\Nb\Lb^{\ab}-\frac12\Lb^{\ab}\Nb\wb_{\ab}-\frac12\psib_{\ua}\Nb\psib^{\ua}\cr 
    &-\frac12\psib^{\ua}\wb_{\ab} \Jb^{\a} (\g_{\ua}\eta)_\a{}^{\ab}+\frac18\Jb^{\ua}(\wb\g_{\ua}\wb)+\Tb_{\mbox{ghosts}},
\end{align}
where $T_{\mbox{ghosts}}$ and $\Tb_{\mbox{ghosts}}$ are the contributions of the RNS ghosts to the stress-energy tensor components and they are equal to the respective contributions in flat spacetime. As was shown in \cite{Chandia:2023eel}, the BRST charge is nilpotent and the BRST currents holomorphic and anti-holomorphic respectively. We now find the BRST transformations of the world-sheet fields and then find the integrated and unintegrated vertex operators.

To construct vertex operators we use canonical commutation relations. The conjugate pairs are $(\L,w), (\Lb,\wb), (\psi,\psi),(\psib,\psib)$and the conjugate of the spacetime super coordinate $Z^M$ is defined as $(-1)^M P_M=\frac{\d S}{\d\p_\s Z^M}$ and is equal to
\begin{align}\label{defP}
    &(-1)^M P_M=\p_\tau Z^N E_{N\ua}E_M{}^{\ua}+\p_\s Z^N B_{NM}-E_M{}^\a d_\a-E_M{}^{\ab}\dd_{\ab}+\frac12\O_{Mab}\left(N^{ab}+\NN^{ab}\right)\cr 
    &+\frac12\O_{Ma'b'}\left(N^{a'b'}+\NN^{a'b'}\right)+\frac12E_M{}^{\ab}\psi^{\ua}(w\eta\g_{\ua})_{\ab}-\frac12E_M{}^\a\psib^{\ua}(\g_{\ua}\eta\wb)_\a\cr
    &-\frac18E_M{}^{\ua}\left((w\g_{\ua}w)+(\wb\g_{\ua}\wb)\right),
\end{align}
and the canonical commutation relations are given by
\begin{align}\label{CANCOM}
    &[\L^\a(\s),w_\b(\s')]=\d^\a_\b\d(\s-\s'),\quad [\Lb^{\ab}(\s),\wb_{\bb}(\s')]=\d^{\ab}_{\bb}\d(\s-\s'),\cr 
    &[\psi^{\ua}(\s),\psi^{\ub}(\s')]=[\psib^{\ua}(\s),\psib^{\ub}(\s')]=\eta^{\ua\ub}\d(\s-\s'),\cr 
    &[Z^M(\s),P_N(\s')]=\d^M_N\d(\s-\s').
\end{align}
Note that we can express $d_\a, \dd_{\ab}$ and $d_{\ua}\equiv\p_\tau Z^M E_{M\ua}$ in terms of canonical variables as
\begin{align}
    &d_\a=(-1)^{M+1}E_\a{}^M P_M+\p_\s Z^M B_{M\a}+\frac12\O_{\a\ua\ub}\left(N^{\ua\ub}+\NN^{\ua\ub}\right)-\frac12\psib^{\ua}(\g_{\ua}\eta\wb)_\a,\cr 
    &\dd_{\ab}=(-1)^{M+1}E_{\ab}{}^M P_M+\p_\s Z^M B_{M\ab}+\frac12\O_{\ab\ua\ub}\left(N^{\ua\ub}+\NN^{\ua\ub}\right)+\frac12\psi^{\ua}(w\eta\g_{\ua})_{\ab},\cr 
    &d_{\ua}=(-1)^ME_{\ua}{}^MP_M-\frac12\O_{\ua\ub\uc}\left(N^{\ub\uc}+\NN^{\ub\uc}\right)+\frac18\left((w\g_{\ua}w)+(\wb\g_{\ua}\wb)\right).
\end{align}


Consider the integrated vertex operator. The most general conformal weight $(1,1)$ has the form
\begin{align}\label{Vads}
    &V=J^\a \Jb^{\ab} A_{\a\ab}+J^{\ua}\Jb^{\ub}A_{\ua\ub}+d_\a\dd_{\ab}P^{\a\ab}+\psi^{\ua}\psi^{\ub}\psib^{\uc}\psib^{\ud}S_{\ua\ub\uc\ud}+\L^\a w_\b\Lb^{\ab}\wb_{\bb}S_{\a\ab}{}^{\b\bb}\cr 
    &+\psi^{\ua}w_\a\psib^{\ub}\wb_{\ab}U_{\ua\ub}{}^{\a\ab}+J^\a\Jb^{\ua}A_{\a\ua}+J^{\ua}\Jb^{\ab}A_{\ua\ab}+J^\a\dd_{\ab}E_\a{}^{\ab}+d_\a\Jb^{\ab}E_{\ab}{}^\a\cr 
&+J^\a\psib^{\ua}\psib^{\ub}\hO_{\a\ua\ub}+\psi^{\ua}\psi^{\ub}\Jb^{\ab} O_{\ab\ua\ub}+J^\a\Lb^{\ab}\wb_{\bb}\hO_{\a\ab}{}^{\bb}+\L^\a w_\b\Jb^{\ab}O_{\ab\a}{}^\b\cr 
    &+J^\a\psib^{\ua}\wb_{\ab} \Ch_{\ua\a}{}^{\ab}+\psi^{\ua}w_\a\Jb^{\ab} C_{\ua\ab}{}^\a+J^{\ua}\dd_{\ab}E_{\ua}{}^{\ab}+d_\a\Jb^{\ua}E_{\ua}{}^\a\cr 
&+J^{\ua}\psib^{\ub}\psib^{\uc}\hO_{\ua\ub\uc}+\psi^{\ub}\psi^{\uc}\Jb^{\ua}O_{\ua\ub\uc}+J^{\ua}\Lb^{\ab}\wb_{\bb}\hO_{\ua\ab}{}^{\bb}+\L^\a w_\b\Jb^{\ua}O_{\ua\a}{}^\b\cr 
    &+J^{\ua}\psib^{\ub}\wb_{\ab}C_{\ua\ub}{}^{\ab}+\psi^{\ub} w_\a\Jb^{\ua}C_{\ua\ub}{}^\a+d_\a\psib^{\ua}\psib^{\ub}\Dh_{\ua\ub}{}^\a+\psi^{\ua}\psi^{\ub}\dd_{\ab}D_{\ua\ub}{}^{\ab} \cr 
    &+d_\a\Lb^{\ab}\wb_{\bb}\Dh_{\ab}{}^{\bb\a}+\L^\a w_\b \dd_{\ab} D_\a{}^{\b\ab}+d_\a\psib^{\ua}\wb_{\ab}V_{\ua}{}^{\a\ab}+\psi^{\ua}w_\a\dd_{\ab}\Vh_{\ua}{}^{\ab\a}\cr 
    &+\psi^{\ua}\psi^{\ub}\Lb^{\ab}\wb_{\bb} S_{\ua\ub\ab}{}^{\bb}+\L^\a w_\b\psib^{\ua}\psib^{\ub}S_{\ua\ub\a}{}^\b+\psi^{\ua}\psi^{\ub}\psib^{\uc}\wb_{\ab}Y_{\ua\ub\uc}{}^{\ab}\cr  
    &+\psi^{\uc}w_\a\psib^{\ua}\psib^{\ub}\Yh_{\ua\ub\uc}{}^\a+\L^\a w_\b\psib^{\ua}\wb_{\ab}Y_{\ua\a}{}^{\b\ab}+\psi^{\ua}w_\a\Lb^{\ab}\wb_{\bb}\Yh_{\ua\ab}{}^{\bb\a}.
\end{align}
We now compute the action of $\g G_L$ and $\gb G_R$ on the integrated vertex $V$ that are defined as
\begin{align}
    \left(\oint \g G_L\right)V=\oint d\s'[\g G_L(\s'),V(\s)],\quad \left(\oint \gb G_R\right)V=\oint d\s'[\gb G_R(\s'),V(\s)],
\end{align}
where we use the commutators shown in the Appendix. 

To compute $\left(\oint \g G_L\right)V$ we proceed as follows. We first note that the integrated vertex operator (\ref{Vads}) can be written as
\begin{align}\label{VL}
    V=\Jb^{\ab}V_{\ab}+\Jb^{\ua}V_{\ua}+\dd_{\ab}V^{\ab}+\psib^{\ua}\psib^{\ub}V_{\ua\ub}+\Lb^{\ab}\wb_{\bb}V_{\ab}{}^{\bb}+\psib^{\ua}\wb_{\ab}V_{\ua}{}^{\ab}, 
\end{align}
where
\begin{align}\label{vertexADS}
    &V_{\ab}=-J^\a A_{\a\ab}+J^{\ua}A_{\ua\ab}-d_\a E_z{\ab}{}^{\a}+\psi^{\ua}\psi^{\ub}O_{\ab\ua\ub}+\L^\a w_\b O_{\ab\a}{}^\b-\psi^{\ua}w_\a C_{\ua\ab}{}^\a,\cr 
    &V_{\ua}=J^\a A_{\a\ua}+J^{\ub}A_{\ub\ua}+d_\a E_{\ua}{}^\a+\psi^{\ub}\psi^{\uc}O_{\ua\ub\uc}+\L^\a w_\b O_{\ua\a}{}^\b+\psi^{\ub} w_\a C_{\ua\ub}{}^\a,\cr
    &V^{\ab}=-J^\a E_\a{}^{\ab}+J^{\ua}E_{\ua}{}^{\ab}-d_\a P^{\a\ab}+\psi^{\ua}\psi^{\ub}D_{\ua\ub}{}^{\ab}+\L^\a w_\b D_\a{}^{\b\ab}-\psi^{\ua}w_\a\Vh_{\ua}{}^{\ab\a} ,\cr
    &V_{\ua\ub}=J^\a \hO_{\a\ua\ub}+J^{\uc}\hO_{\uc\ua\ub}+d_\a\Dh_{\ua\ub}{}^\a+\psi^{\uc}\psi^{\ud}S_{\uc\ud\ua\ub}+\L^\a w_\b S_{\ua\ub\a}{}^\b+\psi^{\uc}w_\a\Yh_{\ua\ub\uc}{}^\a ,\cr &V_{\ab}{}^{\bb}=J^\a\hO_{\a\ab}{}^{\bb}+J^{\ua}\hO_{\ua\ab}{}^{\bb}+d_\a\Dh_{\ab}{}^{\bb\a}+\psi^{\ua}\psi^{\ub}S_{\ua\ub\ab}{}^{\bb}+\L^\a w_\b S_{\a\ab}{}^{\b\bb}+\psi^{\ua}w_\a\Yh_{\ua\ab}{}^{\bb\a} ,\cr 
    &V_{\ua}{}^{\ab}=-J^\a \Ch_{\ua\a}{}^{\ab}+J^{\ub}C_{\ub\ua}{}^{\ab}-d_\a V_{\ua}{}^{\a\ab}+\psi^{\ub}\psi^{\uc}Y_{\ub\uc\ua}{}^{\ab}+\L^\a w_{\b}Y_{\ua\a}{}^{\b\ab}-\psi^{\ub}w_\a U_{\ub\ua}{}^{\a\ab} .
\end{align}
Consider the first term in (\ref{VL}). We obtain
\begin{align}\label{GL1}
    \left(\oint \g G_L\right)\Jb^{\ab}V_{\ab}=\N\left(\g\Jb^{\ab}\left(\L^\a A_{\a\ab}+\psi^{\ua}A_{\ua\ab}-w_\a E_{\ab}{}^\a\right)\right),
\end{align}
provided that the background fields satisfy the equations
\begin{align}\label{GL2}
    &\N_{(\a}A_{\b)\ab}=-\g^{\ua}_{\a\b}A_{\ua\ab},\cr 
    &\N_\a A_{\ua\ab}+\N_{\ua}A_{\a\ab}=-(\g_{\ua})_{\a\b}E_{\ab}{}^\b,\cr
    &O_{\ab\a}{}^\b=-\frac12(\g^{\ua\ub})_\a{}^\b O_{\ab\ua\ub}=\frac14 (\g^{\ua\ub})_\a{}^\b \N_{[\ua}A_{\ub]\ab}.
\end{align}
In the calculation of (\ref{GL1}) the equations derived from (\ref{GL2}) just as in the flat space case. For example, acting with $\N_\a$ one obtains
\begin{align}
    &\N_\a E_{\ab}{}^\b-\frac12\eta^{\b\gb}\N_{\gb}A_{\a\ab}=-\frac14(\g^{\ua\ub})_\a{}^\b \N_{[\ua}A_{\ub]\ab}.
\end{align}
This equation was derived in \cite{Chandia:2017afc} (see also \cite{Chandia:2018nyh,Chandia:2019paj} for a discussion of the generic background geometry). The background field $C_{\ua\ab}{}^\a$ is given by other background fields as
\begin{align}
    C_{\ua\a}{}^\b=-\N_{\ua}E_{\ab}{}^\a+\frac12\eta^{\a\gb}\N_{\gb}A_{\ua\ab}+\frac14(\g_{\ua})^{\a\b}A_{\b\ab}.
\end{align}

A similar calculation can be done for the remaining terms in (\ref{VL}). For the second term in (\ref{VL}) we obtain
\begin{align}\label{GL3}
    \left(\oint \g G_L\right)\Jb^{\ua}V_{\ua}=\N\left(\g\Jb^{\ua}\left(\L^\a A_{\a\ua}-\psi^{\ub}A_{\ub\ua}-w_\a E_{\ua}{}^\a\right)\right),
\end{align}
provided that the background fields satisfy the equations
\begin{align}\label{GL4}
    &\N_{(\a}A_{\b)\ua}=\g^{\ub}_{\a\b}A_{\ub\ua},\cr 
    &\N_\a A_{\ub\ua}-\N_{\ub}A_{\a\ua}=(\g_{\ub})_{\a\b}E_{\ua}{}^\b,\cr
    &O_{\ua\a}{}^\b=-\frac12(\g^{\ub\uc})_\a{}^\b O_{\ua\ub\uc}=\frac14 (\g^{\ub\uc})_\a{}^\b \N_{[\ub}A_{\uc]\ua}.
\end{align}
In the calculation of (\ref{GL3}) the equations derived from (\ref{GL4}) just as in the flat space case. For example, acting with $\N_\a$ one obtains
\begin{align}
    &\N_\a E_{\ua}{}^\b-\frac12\eta^{\b\gb}\N_{\gb}A_{\a\ua}=\frac14(\g^{\ub\uc})_\a{}^\b \N_{[\ub}A_{\uc]\ua}.
\end{align}
The background field $C_{\ua\ub}{}^\a$ is given by other background fields as
\begin{align}
    C_{\ua\ub}{}^\a=-\N_{\ub}E_{\ua}{}^\a-\frac12\eta^{\a\gb}\N_{\gb}A_{\ub\ua}+\frac14(\g_{\ub})^{\a\b}A_{\b\ua}.
\end{align}

For the third term in (\ref{VL}) we obtain
\begin{align}\label{GL5}
    \left(\oint \g G_L\right)\dd_{\ab}V^{\ab}=\N\left(\g\dd_{\ab}\left(\L^\a E_{\a}{}^{\ab}+\psi^{\ua}E_{\ua}{}^{\ab}-w_\a P^{\a\ab}\right)\right),
\end{align}
provided that the background fields satisfy the equations
\begin{align}\label{GL6}
    &\N_{(\a}E_{\b)}{}^{\ab}=-\g^{\ua}_{\a\b}E_{\ua}{}^{\ab},\cr 
    &\N_\a E_{\ua}{}^{\ab}+\N_{\ua}E_{\a}{}^{\ab}=-(\g_{\ua})_{\a\b}P^{\b\ab},\cr
    &D_{\a}{}^{\b\ab}=-\frac12(\g^{\ua\ub})_\a{}^\b D_{\ua\ub}{}^{\ab}=\frac14 (\g^{\ua\ub})_\a{}^\b \N_{[\ua}E_{\ub]}{}^{\ab}.
\end{align}
In the calculation of (\ref{GL3}) the equations derived from (\ref{GL4}) just as in the flat space case. For example, acting with $\N_\a$ one obtains
\begin{align}
    &\N_\a P^{\b\ab}-\frac12\eta^{\b\gb}\N_{\gb}E_\a{}^{\ab}=-\frac14(\g^{\ua\ub})_\a{}^\b \N_{[\ua}E_{\ub]}{}^{\ab}.
\end{align}
The background field $\Vh_{\ua}{}^{\ab\a}$ is given by other background fields as
\begin{align}
    \Vh_{\ua}{}^{\ab\a}=-\N_{\ua}P^{\a\ab}+\frac12\eta^{\a\gb}\N_{\gb}E_{\ua}{}^{\ab}+\frac14\g_{\ua}^{\a\b}E_\b{}^{\ab}.
\end{align}

For the fourth term in (\ref{VL}) we obtain
\begin{align}\label{GL7}
    \left(\oint \g G_L\right)\psib^{\ua}\psib^{\ub}V_{\ua\ub}=\N\left(\g\psib^{\ua}\psib^{\ub}\left(\L^\a \hO_{\a\ua\ub}-\psi^{\uc}\hO_{\uc\ua\ub}-w_\a \Dh_{\ua\ub}{}^\a\right)\right),
\end{align}
provided that the background fields satisfy the equations
\begin{align}\label{GL8}
    &\N_{(\a}\hO_{\b)\ua\ub}=\g^{\uc}_{\a\b}\hO_{\uc\ua\ub},\cr 
    &\N_\a \hO_{\uc\ua\ub}-\N_{\uc}\hO_{\a\ua\ub}=(\g_{\uc})_{\a\b}\Dh_{\ua\ub}{}^\b,\cr
    &S_{\ua\ub\a}{}^\b=-\frac12(\g^{\uc\ud})_\a{}^\b S_{\uc\ud\ua\ub}=\frac14 (\g^{\uc\ud})_\a{}^\b \N_{[\uc}\hO_{\ud]\ua\ub}.
\end{align}
In the calculation of (\ref{GL7}) the equations derived from (\ref{GL8}) just as in the flat space case. For example, acting with $\N_\a$ one obtains
\begin{align}
    &\N_\a \Dh_{\ua\ub}{}^\b-\frac12\eta^{\b\gb}\N_{\gb}\hO_{\a\ua\ub}=\frac14(\g^{\uc\ud})_\a{}^\b \N_{[\uc}\hO_{\ud]\ua\ub}.
\end{align}
The background field $\Yh_{\ua\ub\uc}{}^\a$ is given by other background fields as
\begin{align}
    \Yh_{\ua\ub\uc}{}^\a=-\N_{\uc}\Dh_{\ua\ub}{}^\a
    -\frac12\eta^{\a\gb}\N_{\gb}\hO_{\uc\ua\ub}-\frac14(\g_{\uc})^{\a\b}\hO_{\b\ua\ub}.
\end{align}

For the fifth term in (\ref{VL}) we obtain
\begin{align}\label{GL9}
    \left(\oint \g G_L\right)\Lb^{\ab}\wb_{\bb}V_{\ab}{}^{\bb} =\N\left(\g\Lb^{\ab}\wb_{\bb}\left(\L^\a \hO_{\a\ab}{}^{\bb}-\psi^{\ua}\hO_{\ua\ab}{}^{\bb}-w_\a \Dh_{\ab}{}^{\bb\a}\right)\right),
\end{align}
provided that the background fields satisfy the equations
\begin{align}\label{GL10}
    &\N_{(\a}\hO_{\b)\ab}{}^{\bb}=\g^{\ua}_{\a\b}\hO_{\ua\ab}{}^{\bb},\cr 
    &\N_\a \hO_{\ua\ab}{}^{\bb}-\N_{\ua}\hO_{\a\ab}{}^{\bb}=(\g_{\ua})_{\a\b}\Dh_{\ab}{}^{\bb\b},\cr
    &S_{\a\ab}{}^{\b\bb}=-\frac12(\g^{\ua\ub})_\a{}^\b S_{\ua\ub\ab}{}^{\bb}=\frac14 (\g^{\ua\ub})_\a{}^\b \N_{[\ua}\hO_{\ub]\ab}{}^{\bb}.
\end{align}
In the calculation of (\ref{GL9}) the equations derived from (\ref{GL10}) just as in the flat space case. For example, acting with $\N_\a$ one obtains
\begin{align}
    &\N_\a\Dh_{\ab}{}^{\bb\a}-\frac12\eta^{\b\gb}\N_{\gb}\hO_{\a\ab}{}^{\bb}=\frac14(\g^{\ua\ub})_\a{}^\b \N_{[\ua}\hO_{\ub]\ab}{}^{\bb}.
\end{align}
The background field $\Yh_{\ua\ab}{}^{\bb\a}$ is given by other background fields as
\begin{align}
    \Yh_{\ua\ab}{}^{\bb\a}=-\N_{\ua}\Dh_{\ab}{}^{\bb\a}-\frac12\eta^{\a\gb}\N_{\gb}\hO_{\ua\ab}{}^{\bb}+\frac14(\g_{\ua})^{\a\b}\hO_{\b\ab}{}^{\bb}.
\end{align}

And finally for the sixth term in (\ref{VL}) we obtain
\begin{align}\label{GL11}
    \left(\oint \g G_L\right)\psib^{\ua}\wb_{\ab}V_{\ua}{}^{\ab}=\N\left(\g\psib^{\ua}\wb_{\ab}\left(\L^\a \Ch_{\ua\a}{}^{\ab}+\psi^{\ub}C_{\ub\ua}{}^{\ab}-w_\a V_{\ua}{}^{\a\ab}\right)\right),
\end{align}
provided that the background fields satisfy the equations
\begin{align}\label{GL12}
    &\N_{(\a}\Ch_{\ua\b)}{}^{\ab}=-\g^{\ub}_{\a\b}C_{\ub\ua}{}^{\ab},\cr 
    &\N_\a C_{\ub\ua}{}^{\ab}+\N_{\ub}\Ch_{\ua\a}{}^{\ab}=-(\g_{\ub})_{\a\b}V_{\ua}{}^{\b\ab},\cr
    &Y_{\ua\a}{}^{\b\ab}=-\frac12(\g^{\ub\uc})_\a{}^\b Y_{\ub\uc\ua}{}^{\ab}=\frac14 (\g^{\ub\uc})_\a{}^\b \N_{[\ub}C_{\uc]\ua}{}^{\ab}.
\end{align}
In the calculation of (\ref{GL11}) the equations derived from (\ref{GL12}) just as in the flat space case. For example, acting with $\N_\a$ one obtains
\begin{align}
    &\N_\a V_{\ua}{}^{\b\ab}-\frac12\eta^{\b\gb}\N_{\gb}\Ch_{\ua\a}{}^{\ab}=-\frac14(\g^{\ub\uc})_\a{}^\b \N_{[\ub}C_{\uc]\ua}{}^{\ab}.
\end{align}
The background field $U_{\ub\ua}{}^{\a\ab}$ is given by other background fields as
\begin{align}
    U_{\ub\ua}{}^{\a\ab}=-\N_{\ub}V_{\ua}{}^{\a\ab}+\frac12\eta^{\a\gb}\N_{\gb}C_{\ub\ua}{}^{\ab}+\frac14(\g_{\ub})^{\a\b}\Ch_{\ua\b}{}^{\ab}.
\end{align}

In summary, we have obtained 
\begin{align}
     &\left(\oint \g G_L\right)V=\p (\g w_L),
\end{align}
where
\begin{align}\label{wLis}
    w_L&=\Jb^{\ab}\left(\L^\a A_{\a\ab}+\psi^{\ua}A_{\ua\ab}-w_\a E_{\ab}{}^\a\right)+\Jb^{\ua}\left(\L^\a A_{\a\ua}-\psi^{\ub}A_{\ub\ua}-w_\a E_{\ua}{}^\a\right)\cr
     &+\dd_{\ab}\left(\L^\a E_{\a}{}^{\ab}+\psi^{\ua}E_{\ua\ab}-w_\a P^{\a\ab}\right)+\psib^{\ua}\psib^{\ub}\left(\L^\a \hO_{\a\ua\ub}-\psi^{\uc}\hO_{\uc\ua\ub}-w_\a \Dh_{\ua\ub}{}^\a\right)\cr 
     &+\Lb^{\ab}\wb_{\bb}\left(\L^\a \hO_{\a\ab}{}^{\bb}-\psi^{\ua}\hO_{\ua\ab}{}^{\bb}-w_\a \Dh_{\ab}{}^{\bb\a}\right)+\psib^{\ua}\wb_{\ab}\left(\L^\a \Ch_{\ua\a}{}^{\ab}+\psi^{\ub}C_{\ub\ua}{}^{\ab}-w_\a V_{\ua}{}^{\a\ab}\right),
\end{align}
which resembles the result of the flat space-time background case.

For the calculation of $\left(\oint \g G_R\right)V$, we proceed in a similar way. That is, we recognize that the integrated vertex operator (\ref{Vads}) can be written as
\begin{align}\label{VR}
    V=J^{\a}\Vh_{\a}+J^{\ua}\Vh_{\ua}+d_{\a}\Vh^{\a}+\psi^{\ua}\psi^{\ub}\Vh_{\ua\ub}+\L^{\a}w_{\b}\Vh_{\a}{}^{\b}+\psi^{\ua}w_{\a}\Vh_{\ua}{}^{\a}, 
\end{align}
where
\begin{align}\label{vertexADSP}
    &\Vh_{\a}=\Jb^{\ab} A_{\a\ab}+\Jb^{\ua}A_{\a\ua}+\dd_{\ab} E_{\a}{}^{\ab}+\psib^{\ua}\psib^{\ub}\hO_{\a\ua\ub}+\Lb^{\ab} \wb_{\bb} \hO_{\a\ab}{}^{\bb}+\psib^{\ua}\wb_{\ab} \Ch_{\ua\a}{}^{\ab},\cr 
    &\Vh_{\ua}=\Jb^{\ab} A_{\ua\ab}+\Jb^{\ub}A_{\ua\ub}+\dd_{\ab} E_{\ua}{}^{\ab}+\psib^{\ub}\psib^{\uc}\hO_{\ua\ub\uc}+\Lb^{\ab} \wb_{\bb} \hO_{\ua\ab}{}^{\bb}+\psib^{\ub} \wb_{\ab} C_{\ua\ub}{}^{\ab},\cr
    &\Vh^{\a}=\Jb^{\ab} E_{\ab}{}^{\a}+\Jb^{\ua}E_{\ua}{}^{\a}+\dd_{\ab} P^{\a\ab}+\psib^{\ua}\psib^{\ub}\Dh_{\ua\ub}{}^{\a}+\Lb^{\ab} \wb_{\bb} \Dh_{\ab}{}^{\bb\a}+\psib^{\ua}\wb_{\ab} V_{\ua}{}^{\a\ab} ,\cr
    &\Vh_{\ua\ub}=\Jb^{\ab} O_{\ab\ua\ub}+\Jb^{\uc} O_{\uc\ua\ub}+\dd_{\ab} D_{\ua\ub}{}^{\ab}+\psib^{\uc}\psib^{\ud}S_{\ua\ub\uc\ud}+\Lb^{\ab} \wb_{\bb} S_{\ua\ub\ab}{}^{\bb}+\psib^{\uc}\wb_{\ab} Y_{\ua\ub\uc}{}^{\ab} ,\cr 
    &\Vh_{\a}{}^{\b}=\Jb^{\ab} O_{\ab\a}{}^{\b}+\Jb^{\ua} O_{\ua\a}{}^{\b}+\dd_{\ab} D_{\a}{}^{\b\ab}+\psib^{\ua}\psib^{\ub}S_{\ua\ub\a}{}^{\b}+\Lb^{\ab} \wb_{\bb} S_{\a\ab}{}^{\b\bb}+\psib^{\ua}\wb_{\ab} Y_{\ua\a}{}^{\b\ab} ,\cr 
    &\Vh_{\ua}{}^{\a}=\Jb^{\ab} C_{\ua\ab}{}^{\a}+\Jb^{\ub}C_{\ub\ua}{}^{\a}+\dd_{\ab} \Vh_{\ua}{}^{\ab\a}+\psib^{\ub}\psib^{\uc}\Yh_{\ub\uc\ua}{}^{\a}+\Lb^{\ab} \wb_{\bb}\Yh_{\ua\ab}{}^{\bb\a}+\psib^{\ub}\wb_{\ab} U_{\ua\ub}{}^{\a\ab} .
\end{align}

Consider the first term in (\ref{VR}). We obtain
\begin{align}\label{GR1}
    \left(\oint \gb G_R\right)J^{\a}\Vh_{\a}=\Nb\left(\gb J^{\a}\left(-\Lb^{\b} A_{\a\ab}+\psib^{\ua}A_{\a\ua}+\wb_{\ab} E_{\a}{}^{\ab}\right)\right),
\end{align}
provided that the background fields satisfy the equations
\begin{align}\label{GR2}
    &\N_{(\ab}A_{\a\bb)}=\g^{\ua}_{\ab\bb}A_{\a\ua},\cr 
    &\N_{\ab} A_{\a\ua}-\N_{\ua}A_{\a\ab}=(\g_{\ua})_{\ab\bb}E_{\a}{}^{\bb},\cr
    &\hO_{\a\ab}{}^{\bb}=-\frac12(\g^{\ua\ub})_{\ab}{}^{\bb} \hO_{\a\ua\ub}=\frac14 (\g^{\ua\ub})_{\ab}{}^{\bb} \N_{[\ua}A_{\a\ub]}.
\end{align}
In the calculation of (\ref{GR1}) the equations derived from (\ref{GR2}) just as in the flat space case. For example, acting with $\N_{\ab}$ one obtains
\begin{align}
    &\N_{\ab} E_{\a}{}^{\bb}+\frac12\eta^{\g\bb}\N_{\g}A_{\a\ab}=\frac14(\g^{\ua\ub})_{\ab}{}^{\bb} \N_{[\ua}A_{\a\ub]}.
\end{align}
The background field $\Ch_{\ua\a}{}^{\ab}$ is given by other background fields as
\begin{align}
    \Ch_{\ua\a}{}^{\ab}=\N_{\ua}E_{\a}{}^{\ab}+\frac12\eta^{\g\ab}\N_{\g}A_{\a\ua}+\frac14(\g_{\ua})^{\ab\bb}A_{\a\bb}.
\end{align}

A similar calculation can be done for the remaining terms in (\ref{VR}). For the second term in (\ref{VR}) we obtain
\begin{align}\label{GR3}
    \left(\oint \g G_R\right) J^{\ua}\Vh_{\ua}=\Nb\left(\gb J^{\ua}\left(\Lb^{\ab} A_{\ua\a}-\psib^{\ub}A_{\ua\ub}-\wb_{\ab} E_{\ua}{}^{\ab}\right)\right),
\end{align}
provided that the background fields satisfy the equations
\begin{align}\label{GR4}
    &\N_{(\ab}A_{\ua\bb)}=\g^{\ub}_{\ab\bb}A_{\ua\ub},\cr 
    &\N_{\ab} A_{\ua\ub}-\N_{\ub}A_{\ua\ab}=(\g_{\ub})_{\ab\bb}E_{\ua}{}^{\bb},\cr
    &\hO_{\ua\ab}{}^{\bb}=-\frac12(\g^{\ub\uc})_{\ab}{}^{\bb} \hO_{\ua\ub\uc}=\frac14 (\g^{\ub\uc})_{\ab}{}^{\bb} \N_{[\ub}A_{\ua\uc]}.
\end{align}
In the calculation of (\ref{GR3}) the equations derived from (\ref{GL4}) just as in the flat space case. For example, acting with $\N_{\ab}$ one obtains
\begin{align}
    &\N_{\ab} E_{\ua}{}^{\bb}+\frac12\eta^{\g\bb}\N_{\g}A_{\ua\ab}=\frac14(\g^{\ub\uc})_{\ab}{}^{\bb} \N_{[\ub}A_{\ua\uc]}.
\end{align}
The background field $C_{\ua\ub}{}^{\ab}$ is given by other background fields as
\begin{align}
    C_{\ua\ub}{}^{\ab}=\N_{\ub}E_{\ua}{}^{\ab}+\frac12\eta^{\g\ab}\N_{\g}A_{\ua\ub}+\frac14(\g_{\ub})^{\ab\bb}A_{\ua\bb}.
\end{align}

For the third term in (\ref{VR}) we obtain
\begin{align}\label{GR5}
    \left(\oint \g G_R\right) d_{\a}\Vh^{\a}=\Nb\left(\gb d_{\a}\left(-\Lb^{\ab} E_{\ab}{}^{\a}-\psib^{\ua}E_{\ua}{}^{\a}+\wb_{\ab} P^{\a\ab}\right)\right),
\end{align}
provided that the background fields satisfy the equations
\begin{align}\label{GR6}
    &\N_{(\ab}E_{\bb)}{}^{\a}=\g^{\ua}_{\ab\bb}E_{\ua}{}^{\a},\cr 
    &\N_{\ab} E_{\ua}{}^{\a}-\N_{\ua}E_{\ab}{}^{\a}=(\g_{\ua})_{\ab\bb}P^{\a\bb},\cr
    &\Dh_{\ab}{}^{\bb\a}=-\frac12(\g^{\ua\ub})_{\ab}{}^{\bb} \Dh_{\ua\ub}{}^{\a}=\frac14 (\g^{\ua\ub})_{\ab}{}^{\bb} \N_{[\ua}E_{\ub]}{}^{\a}.
\end{align}
In the calculation of (\ref{GR3}) the equations derived from (\ref{GL4}) just as in the flat space case. For example, acting with $\N_{\ab}$ one obtains
\begin{align}
    &\N_{\ab} P^{\a\bb}+\frac12\eta^{\g\bb}\N_{\g}E_{\ab}{}^{\a}=\frac14(\g^{\ua\ub})_{\ab}{}^{\bb} \N_{[\ua}E_{\ub]}{}^{\a}.
\end{align}
The background field $V_{\ua}{}^{\a\ab}$ is given by other background fields as
\begin{align}
    V_{\ua}{}^{\a\ab}=\N_{\ua}P^{\a\ab}+\frac12\eta^{\g\ab}\N_{\g}E_{\ua}{}^{\a}+\frac14\g_{\ua}^{\ab\bb}E_{\bb}{}^{\a}.
\end{align}

For the fourth term in (\ref{VR}) we obtain
\begin{align}\label{GR7}
    \left(\oint \gb G_R\right)\psi^{\ua}\psi^{\ub}\Vh_{\ua\ub}=\Nb\left(\gb\psi^{\ua}\psi^{\ub}\left(\Lb^{\ab} O_{\ab\ua\ub}-\psib^{\uc} O_{\uc\ua\ub}-\wb_{\ab} D_{\ua\ub}{}^{\ab}\right)\right),
\end{align}
provided that the background fields satisfy the equations
\begin{align}\label{GR8}
    &\N_{(\ab} O_{\bb)\ua\ub}=\g^{\uc}_{\ab\bb} O_{\uc\ua\ub},\cr 
    &\N_{\ab} O_{\uc\ua\ub}-\N_{\uc} O_{\ab\ua\ub}=(\g_{\uc})_{\ab\bb} D_{\ua\ub}{}^{\bb},\cr
    &S_{\ua\ub\ab}{}^{\bb}=-\frac12(\g^{\uc\ud})_{\ab}{}^{\bb} S_{\ua\ub\uc\ud}=\frac14 (\g^{\uc\ud})_{\ab}{}^{\bb} \N_{[\uc} O_{\ud]\ua\ub}.
\end{align}
In the calculation of (\ref{GR7}) the equations derived from (\ref{GR8}) just as in the flat space case. For example, acting with $\N_{\ab}$ one obtains
\begin{align}
    &\N_{\ab} D_{\ua\ub}{}^{\bb}+\frac12\eta^{\g\bb}\N_{\g} O_{\ab\ua\ub}=\frac14(\g^{\uc\ud})_{\ab}{}^{\bb} \N_{[\uc} O_{\ud]\ua\ub}.
\end{align}
The background field $Y_{\ua\ub\uc}{}^{\ab}$ is given by other background fields as
\begin{align}
    Y_{\ua\ub\uc}{}^{\ab}=\N_{\uc} D_{\ua\ub}{}^{\ab}
    +\frac12\eta^{\g\ab}\N_{\g} O_{\uc\ua\ub}+\frac14(\g_{\uc})^{\ab\bb} O_{\bb\ua\ub}.
\end{align}

For the fifth term in (\ref{VR}) we obtain
\begin{align}\label{GR9}
    \left(\oint \gb G_R\right)\L^{\a} w_{\b}\Vh_{\a}{}^{\b} =\Nb\left(\gb\L^{\a} w_{\b}\left(\Lb^{\ab} O_{\ab\a}{}^{\b}-\psib^{\ua} O_{\ua\a}{}^{\b}-\wb_{\ab} D_{\a}{}^{\b\ab}\right)\right),
\end{align}
provided that the background fields satisfy the equations
\begin{align}\label{GR10}
    &\N_{(\ab} O_{\bb)\a}{}^{\b}=\g^{\ua}_{\ab\bb} O_{\ua\a}{}^{\b},\cr 
    &\N_{\ab} O_{\ua\a}{}^{\b}-\N_{\ua} O_{\ab\a}{}^{\b}=(\g_{\ua})_{\ab\bb} D_{\a}{}^{\b\bb},\cr
    &S_{\a\ab}{}^{\b\bb}=-\frac12(\g^{\ua\ub})_{\ab}{}^{\bb} S_{\ua\ub\a}{}^{\b}=\frac14 (\g^{\ua\ub})_{\ab}{}^{\bb} \N_{[\ua} O_{\ub]\a}{}^{\b}.
\end{align}
In the calculation of (\ref{GR9}) the equations derived from (\ref{GR10}) just as in the flat space case. For example, acting with $\N_{\ab}$ one obtains
\begin{align}
    &\N_{\ab} D_{\a}{}^{\b\bb}+\frac12\eta^{\g\bb}\N_{\g} O_{\ab\a}{}^{\b}=\frac14(\g^{\ua\ub})_{\ab}{}^{\bb} \N_{[\ua} O_{\ub]\a}{}^{\b}.
\end{align}
The background field $Y_{\ua\a}{}^{\b\ab}$ is given by other background fields as
\begin{align}
    Y_{\ua\a}{}^{\b\ab}=\N_{\ua}D_{\a}{}^{\b\ab}+\frac12\eta^{\g\ab}\N_{\g}O_{\ua\a}{}^{\b}+\frac14(\g_{\ua})^{\ab\bb}O_{\bb\a}{}^{\b}.
\end{align}

And finally for the sixth term in (\ref{VR}) we obtain
\begin{align}\label{GR11}
    \left(\oint \gb G_R\right)\psi^{\ua} w_{\a}\Vh_{\ua}{}^{\a}=\Nb\left(\gb\psi^{\ua} w_{\a}\left(-\Lb^{\ab} C_{\ua\ab}{}^{\a}+\psib^{\ub}C_{\ub\ua}{}^{\a}+\wb_{\ab} \Vh_{\ua}{}^{\ab\a}\right)\right),
\end{align}
provided that the background fields satisfy the equations
\begin{align}\label{GR12}
    &\N_{(\ab} C_{\ua\bb)}{}^{\a}=\g^{\ub}_{\ab\bb}C_{\ub\ua}{}^{\a},\cr 
    &\N_{\ab} C_{\ub\ua}{}^{\a}-\N_{\ub} C_{\ua\ab}{}^{\a}=(\g_{\ub})_{\ab\bb}\Vh_{\ua}{}^{\bb\a},\cr
    &\Yh_{\ua\ab}{}^{\bb\a}=-\frac12(\g^{\ub\uc})_{\ab}{}^{\bb} \Yh_{\ub\uc\ua}{}^{\a}=\frac14 (\g^{\ub\uc})_{\ab}{}^{\bb} \N_{[\ub}C_{\uc]\ua}{}^{\a}.
\end{align} 
In the calculation of (\ref{GL11}) the equations derived from (\ref{GR12}) just as in the flat space case. For example acting with $\N_{\ab}$ one obtains
\begin{align}
    &\N_{\ab} \Vh_{\ua}{}^{\bb\a}+\frac12\eta^{\g\bb}\N_{\g} C_{\ua\ab}{}^{\a}=\frac14(\g^{\ub\uc})_{\ab}{}^{\bb} \N_{[\ub}C_{\uc]\ua}{}^{\a}.
\end{align}
The background field $U_{\ua\ub}{}^{\a\ab}$ is given by other background fields as
\begin{align}
    U_{\ua\ub}{}^{\a\ab}=\N_{\ub}\Vh_{\ua}{}^{\ab\a}+\frac12\eta^{\g\ab}\N_{\g}C_{\ub\ua}{}^{\a}+\frac14(\g_{\ub})^{\ab\bb} C_{\ua\bb}{}^{\a}.
\end{align}

Then, we have obtained the following
\begin{align}
     &\left(\oint \g G_R\right)V=\pb (\gb w_R), 
\end{align}
where
\begin{align}\label{wRis}
    w_R&=J^{\a}\left(-\Lb^{\ab} A_{\a\ab}+\psib^{\ua}A_{\a\ua}+\wb_{\ab} E_{\a}{}^{\ab}\right)+ J^{\ua}\left(\Lb^{\ab} A_{\ua\ab}-\psib^{\ub}A_{\ua\ub}-\wb_{\ab} E_{\ua}{}^{\ab}\right)\cr
     &+d_{\a}\left(-\Lb^{\ab} E_{\ab}{}^{\a}+\psib^{\ua}E_{\ua}{}^{\a}+\wb_{\ab} P^{\a\ab}\right)+\psi^{\ua}\psi^{\ub}\left(\Lb^{\ab} O_{\ab\ua\ub}-\psib^{\uc} O_{\uc\ua\ub}-\wb_{\ab} D_{\ua\ub}{}^{\ab}\right)\cr 
     &+\L^{\a} w_{\b}\left(\Lb^{\ab} O_{\ab\a}{}^{\b}-\psib^{\ua} O_{\ua\a}{}^{\b}-\wb_{\ab} D_{\a}{}^{\b\ab}\right)+\psi^{\ua} w_{\a}\left(-\Lb^{\ab} C_{\ua\ab}{}^{\a}+\psib^{\ub}C_{\ub\ua}{}^{\a}+\wb_{\ab} \Vh_{\ua}{}^{\ab\a}\right),
\end{align}
which again resembles the result of the flat space-time background case.

We have completed the calculation of $\oint\left(\g G_L+\gb G_R\right)V$ obtaining
\begin{align}\label{res1}
    \oint\left(\g G_L+\gb G_R\right)V=\p(\g w_L)+\pb(\gb w_R),
\end{align}
where $w_L$ is given in (\ref{wLis}) and $w_R$ is given in (\ref{wRis}).

It remains to determine $\oint\left(c T+\overline c \overline T\right)V$. Using the results of \cite{Chandia:2023eel} we check that
\begin{align}
    \left(\oint c T\right)V=\p(cV),\quad \left(\oint \overline c \Tb\right)V=\pb(\overline cV)
\end{align}
to finally obtain
\begin{align}
    QV=\pb W-\p\Wb,
\end{align}
where
\begin{align}
    W=\overline c V+\gb w_R,\quad \Wb=-c V-\g w_L,
\end{align}
just like in the flat space-time case (\ref{Wflat}).

The last step in our construction is to determine the unintegrated vertex operator, $U$, that satisfies
\begin{align}
    QW=\p U,\quad Q\Wb=\pb U. 
\end{align}
Consider $W$. Using Since $T$ and $\Tb$ act on world-sheet fields creating a world-sheet derivative on it, we obtain
\begin{align}\label{QWads}
    &QW=\p\left(c\overline c V+c\gb w_R-\overline c\g w_L \right)+\gb^2 V\cr 
    &+\gb\left(\oint\gb G_R\right)w_R+\gb\left(\oint\g G_L\right)w_R.
\end{align}
It will be shown that $\left(\oint\g G_L\right)w_R=\p \g u$ for some $u$ which will be given below and that $\left(\oint\gb G_R\right)w_R=-\gb V$ so that the term $\gb^2 V$ in (\ref{QWads}) is canceled. 

Consider $\left(\oint\g G_L\right)w_R$. It is useful to write $w_R$ as
\begin{align}
    w_R&=\Lb^{\ab}\left(-J^\a A_{\a\ab}+J^{\ua}A_{\ua\a}-d_\a E_{\ab}{}^\a +\psi^{\ua}\psi^{\ub}O_{\ab\ua\ub}+\L^\a w_\b O_{\ab\a}{}^\b-\psi^{\ua} w_\a C_{\ua\ab}{}^\a\right) \cr 
    &+\psib^{\ua}\left(-J^\a A_{\a\ua}-J^{\ub}A_{\ub\ua}-d_\a E_{\ua}{}^\a-\psi^{\ub}\psi^{\uc}O_{\ua\ub\uc}-\L^\a w_\b O_{\ua\a}{}^\b-\psi^{\ub}w_\a C_{\ua\ub}{}^\aº\right)\cr 
    &+\wb_{\ab}\left(J^\a E_\a{}^{\ab}-J^{\ua}E_{\ua}{}^{\ab}+d_\a P^{\a\ab}-\psi^{\ua}\psi^{\ub} D_{\ua\ub}{}^{\ab}-\L^\a w_\b D_\a{}^{\b\ab}+\psi^{\ua}w_\a\Vh_{\ua}{}^{\ab\a}\right).
\end{align}
Using the equations for the background superfields and the commutators of the appendix one obtains the following
\begin{align}
    \left(\oint\g G_L\right)w_R=\p(\g u),
\end{align}
where
\begin{align}
    u&=-\L^\a\Lb^{\ab} A_{\a\ab}+\L^\a\psib^{\ua}A_{\a\ua}-\psi^{\ua}\Lb^{\ab}A_{\ua\ab}+\L^\a\wb_{\ab} E_\a{}^{\ab}-w_\a\Lb^{\ab}E_{\ab}{}^\a\cr 
    &+\psi^{\ua}\psib^{\ub}A_{\ua\ub}+\psi^{\ua}\wb_{\ab}E_{\ua}{}^{\ab}-w_\a\psib^{\ua}E_{\ua}{}^\a-w_\a\wb_{\ab} P^{\a\ab},
\end{align}
which takes the same form as the flat spacetime background case. 

Consider now $\left(\oint\gb G_R\right)w_R$. Using the form (\ref{wRis}) and using the equations of the backgound fields we obtain
\begin{align}
    \left(\oint\gb G_R\right)w_R=-\gb V.
\end{align}
Therefore, the unintegrated vertex operator $U$ that satisfies $QW=\p U$ is given by
\begin{align}
    U=c\overline c V+c\gb w_R-\overline c\g w_L+\g\gb u,
\end{align}
which has the form of the unintegrated vertex operator in flat spacetime background (\ref{Uflat}).


\section{Concluding remarks}

In this paper, integrated and unintegrated vertex operators have been constructed in flat spacetime and in the background with the geometry of $AdS_5\times S^5$. It would be interesting to perform a similar construction in a generic background geometry as it is done in \cite{Chandia:2018nyh,Chandia:2019paj} in the pure spinor formalism. A more ambitious goals would be to study scattering amplitudes and to compactify superstrings as it is done in \cite{Chandia:2022uyy,Chandia:2024rze} for pure spinor strings.






\appendix 

\section{Commutators}
In this section, we list the relevant equal world-sheet times Poisson brackets used in the calculations done in the B-RNS-GSS stype IIB superstring in a $AdS_5\times S^5$ background. Using (\ref{CANCOM}) we obtain (all commutators contain a $\d(\s-\s')$ factor where is not shown explicitly)

\begin{align}
    &[d_\a,\L^\b]=-\frac14\O_{\a\ua\ub}(\L\g^{\ua\ub})^\b,\cr
    &[\dd_{\ab},\L^\b]=-\frac14\O_{\ab\ua\ub}(\L\g^{\ua\ub})^\b-\psi^{\ua}T_{\ua\ab}{}^\b,\cr 
    &[d_{\ua},\L^\b]=\frac14\O_{\ua\ub\uc}(\L\g^{\ub\uc})^\b-\frac14(\g_{\ua}w)^\b,\cr 
    &[J^\a,\L^\b]=[\Jb^\a,\L^\b]=\frac12\eta^{\a\ab}\left(-\frac14\O_{\ab\ua\ub}(\L\g^{\ua\ub})^\b-\psi^{\ua}T_{\ua\ab}{}^\b\right),\cr
    &[\Jb^{\ab},\L^\b]=[J^{\ab},\L^\b]=-\frac12\eta^{\a\ab}\left(-\frac14\O_{\a\ua\ub}(\L\g^{\ua\ub})^\b\right),
\end{align}

\begin{align}
    &[d_\a,\Lb^{\bb}]=-\frac14\O_{\a\ua\ub}(\Lb\g^{\ua\ub})^{\bb}-\psib^{\ua}T_{\ua\a}{}^{\bb},\cr 
    &[\dd_{\ab},\Lb^{\bb}]=-\frac14\O_{\ab\ua\ub}(\Lb\g^{\ua\ub})^{\bb},\cr 
    &[d_{\ua},\Lb^{\bb}]=\frac14\O_{\ua\ub\uc}(\Lb\g^{\ub\uc})^{\bb}-\frac14(\g_{\ua}\wb)^{\bb},\cr 
    &[J^\a,\Lb^{\bb}]=[\Jb^\a,\Lb^{\bb}]=\frac12\eta^{\a\ab}\left(-\frac14\O_{\ab\ua\ub}(\Lb\g^{\ua\ub})^{\bb}\right),\cr 
    &[\Jb^{\ab},\Lb^{\bb}]=[J^{\ab},\Lb^{\bb}]=-\frac12\eta^{\a\ab}\left(-\frac14\O_{\a\ua\ub}(\Lb\g^{\ua\ub})^{\bb}-\psib^{\ua}T_{\ua\a}{}^{\bb}\right),
\end{align}

\begin{align}
    &[d_\a,w_\b]=\frac14\O_{\a\ua\ub}(\g^{\ua\ub}w)_\b,\cr 
    &[\dd_{\ab},w_\b]=\frac14\O_{\ab\ua\ub}(\g^{\ua\ub}w)_\b,\cr 
    &[d_{\ua},w_\b]=-\frac14\O_{\ua\ub\uc}(\g^{\ub\uc}w)_\b,\cr 
    &[J^\a,w_\b]=[\Jb^\a,w_\b]=\frac12\eta^{\a\ab}\left(\frac14\O_{\ab\ua\ub}(\g^{\ua\ub}w)_\b\right),\cr 
    &[\Jb^{\ab},w_\b]=[J^{\ab},w_\b]=-\frac12\eta^{\a\ab}\left(\frac14\O_{\a\ua\ub}(\g^{\ua\ub}w)_\b\right),
\end{align}

\begin{align}
    &[d_\a,\wb_{\bb}]=\frac14\O_{\a\ua\ub}(\g^{\ua\ub}\wb)_{\bb},\cr
    &[\dd_{\ab},\wb_{\bb}]=\frac14\O_{\ab\ua\ub}(\g^{\ua\ub}\wb)_{\bb},\cr 
    &[d_{\ua},\wb_{\bb}]=-\frac14\O_{\ua\ua\ub}(\g^{\ua\ub}\wb)_{\bb},\cr 
    &[J^\a,\wb_{\bb}]=[\Jb^\a,\wb_{\bb}]=\frac12\eta^{\a\ab}\left(\frac14\O_{\ab\ua\ub}(\g^{\ua\ub}\wb)_{\bb}\right),\cr 
    &[\Jb^{\ab},\wb_{\bb}]=[J^{\ab},\wb_{\bb}]=-\frac12\eta^{\a\ab}\left(\frac14\O_{\a\ua\ub}(\g^{\ua\ub}\wb)_{\bb}\right),
\end{align}

\begin{align}
    &[d_\a,\psi_{\ua}]=\O_{\a\ua\ub}\psi^{\ub},\cr 
    &[\dd_{\ab},\psi_{\ua}]=\O_{\ab\ua\ub}\psi^{\ub}+\frac12(w\eta\g_{\ua})_{\ab},\cr 
    &[d_{\ua},\psi_{\ub}]=-\O_{\ua\ub\uc}\psi^{\uc},\cr 
    &[J^\a,\psi_{\ua}]=[\Jb^\a,\psi_{\ua}]=\frac12\eta^{\a\ab}\left(\O_{\ab\ua\ub}\psi^{\ub}+\frac12(w\eta\g_{\ua})_{\ab}\right),\cr 
    &[\Jb^{\ab},\psi_{\ua}]=[J^{\ab},\psi_{\ua}]=-\frac12\eta^{\a\ab}\left(\O_{\a\ua\ub}\psi^{\ub}\right),
\end{align}

\begin{align}
    &[d_\a,\psib_{\ua}]=\O_{\a\ua\ub}\psib^{\ub}-\frac12(\g_{\ua}\eta\wb)_\a,\cr 
    &[\dd_{\ab},\psib_{\ua}]=\O_{\ab\ua\ub}\psib^{\ub},\cr 
    &[d_{\ua},\psib_{\ub}]=-\O_{\ua\ub\uc}\psib^{\uc},\cr
    &[J^\a,\psib_{\ua}]=[\Jb^\a,\psib_{\ua}]=\frac12\eta^{\a\ab}\left(\O_{\ab\ua\ub}\psib^{\ub}\right) ,\cr 
    &[\Jb^{\ab},\psib_{\ua}]=[J^{\ab},\psib_{\ua}]=-\frac12\eta^{\a\ab}\left(\O_{\a\ua\ub}\psib^{\ub}-\frac12(\g_{\ua}\eta\wb)_\a\right) ,
\end{align}

\begin{align}
    &[d_\a(\s),d_\b(\s')]=E_\a{}^M(\s)B_{M\b}(\s')\frac{\p}{\p\s'}\d(\s-\s')+E_\b{}^M(\s')B_{M\a}(\s)\frac{\p}{\p\s}\d(\s-\s')\cr
    &+\left((-1)^M\p_\s Z^N E_{(\a}{}^M\p_{\b)}B_{MN}-\p_\tau Z^M E_M{}^{\ua}(\g_{\ua})_{\a\b}+\O_{(\a\b)}{}^\g d_\g +\frac18\g^{\ua}_{\a\b}(w\g_{\ua}w)\right)\d(\s-\s')
\end{align}

\begin{align}
    &[d_\a(\s),d_{\ab}(\s')]=E_\a{}^M(\s)B_{M\ab}(\s')\frac{\p}{\p\s'}\d(\s-\s')+E_{\ab}{}^M(\s')B_{M\a}(\s)\frac{\p}{\p\s}\d(\s-\s')\cr
    &+\left((-1)^M\p_\s Z^N E_{(\a}{}^M\p_{\ab)}B_{MN}+\O_{\ab\a}{}^\b d_\b+\O_{\a\ab}{}^{\bb}\dd_{\bb} +\frac12R_{\a\ab\ua\ub}\left(N^{\ua\ub}+\NN^{\ua\ub}\right)\right)\d(\s-\s')
\end{align}

\begin{align}
    &[d_{\ab}(\s),d_{\bb}(\s')]=E_{\ab}{}^M(\s)B_{M\bb}(\s')\frac{\p}{\p\s'}\d(\s-\s')+E_{\bb}{}^M(\s')B_{M\ab}(\s)\frac{\p}{\p\s}\d(\s-\s')\cr
    &+\left((-1)^M\p_\s Z^N E_{(\ab}{}^M\p_{\bb)}B_{MN}-\p_\tau Z^M E_M{}^{\ua}(\g_{\ua})_{\ab\bb}+\O_{(\ab\bb)}{}^{\gb} \dd_{\gb} +\frac18\g^{\ua}_{\ab\bb}(\wb\g_{\ua}\wb)\right)\d(\s-\s')
\end{align}
\begin{align}
    &[d_{\ua}(\s),d_\a(\s')]=-E_{\ua}{}^M(\s) B_{M\a}(\s')\frac{\p}{\p\s'}\d(\s-\s'),\cr
    &+\left((-1)^N\p_\s Z^N E_\a{}^M \p_{\ua} B_{MN}-(-1)^{M+N}\p_\s Z^N E_{\ua}{}^M \p_\a B_{MN}-\p_\tau Z^M E_M{}^{\ub}\O_{\a\ua\ub}\right.\cr
    &-\left.\O_{\ua\a}{}^\b d_\b +T_{\ua\a}{}^{\bb}\dd_{\bb}-\frac12 T_{\ua\a}{}^{\bb}\psi^{\ub}(w\eta\g_{\ub})_{\bb} \right)\d(\s-\s')
\end{align}

\begin{align}
    &[d_{\ua}(\s),d_{\ab}(\s')]=-E_{\ua}{}^M(\s) B_{M\ab}(\s')\frac{\p}{\p\s'}\d(\s-\s'),\cr
    &+\left((-1)^N\p_\s Z^N E_{\ab}{}^M \p_{\ua} B_{MN}-(-1)^{M+N}\p_\s Z^N E_{\ua}{}^M \p_{\ab} B_{MN}- \p_\tau Z^M E_M{}^{\ub}\O_{\ab\ua\ub}\right.\cr
    &-\left.\O_{\ua\ab}{}^{\bb} \dd_{\bb} +T_{\ua\ab}{}^{\b} d_{\b}+\frac12 T_{\ua\ab}{}^{\b}\psib^{\ub}(\g_{\ub}\eta\wb)_{\b} \right)\d(\s-\s')
\end{align}

\begin{align}
    &[d_{\ua}(\s),d_{\ub}(\s')]=\left(\p_\s Z^M E_{[a}{}^N\p_{b]}B_{NM}-\p_\tau Z^M E_M{}^{\uc}\O_{[\ua\ub]\uc}+\frac12 R_{\ua\ub\uc\ud}\left(N^{\uc\ud}+\NN^{\uc\ud}\right)  \right) \d(\s-\s')
\end{align}





{\small
\bibliographystyle{abe}
\bibliography{mybib}{}
}

\end{document}